\documentclass[aps,showpacs]{revtex4}
\usepackage{graphicx}

\begin{document}

\title[Chaos in compact binaries]{Ruling out chaos in compact binary systems}
\author{J.~D.~Schnittman}
\email{schnittm@mit.edu}
\author{F.~A.~Rasio}
\email{rasio@mit.edu}
\affiliation{Dept.\ of Physics, Massachusetts Institute of Technology\\
Cambridge, MA 02139}
\date{\today}
\begin{abstract}
We investigate the orbits of compact binary systems during the
final inspiral period before coalescence by integrating
numerically the second-order post-Newtonian equations of motion.
We include spin-orbit and spin-spin coupling terms, which,
according to a recent study by Levin [J. Levin, Phys. Rev. Lett.
\textbf{84}, 3515 (2000)], may cause the orbits to become chaotic.
To examine this claim, we study the divergence of initially nearby
phase-space trajectories and attempt to measure the Lyapunov
exponent $\gamma$. Even for systems with maximally spinning
objects and large spin-orbit misalignment angles, we find
\textit{no chaotic behavior}. For all the systems we consider, we
can place a strict lower limit on the divergence time $t_{\rm L}
\equiv 1/\gamma$ that is many times greater than the typical
inspiral time, suggesting that chaos should not adversely affect
the detection of inspiral events by upcoming gravitational-wave
detectors.
\end{abstract}

\pacs{04.30.Db,04.25.Nx,95.30.Sf,05.45.Jn}

\maketitle

For the current generation of ground-based gravitational-wave (GW)
detectors, such as LIGO, one of the most promising sources is the
final inspiral of two compact stars (black holes or neutron stars
in a relativistic binary orbit) \cite{thorne, cutler}. To detect
such an event successfully, one must be able to match theoretical
GW templates to experimental data containing a great deal of
instrumental noise. This ``matched filtering'' technique has the
potential of greatly increasing the effective signal-to-noise
ratio for the detector \cite{thorne, owen}. Even to specify
quantitative \textit{upper limits} on inspiral events (a major
objective of early LIGO runs), it is critical to have strong
confidence in the physical accuracy of the templates \cite{allen}.

Recent work has suggested that the orbits of two rapidly spinning
compact objects may be chaotic \cite{levinprl, levingrqc}. The
presence of chaos on the inspiral timescale (typically
$\sim100\,$s as the frequency sweeps up from $\sim 10$ to
$10^3\,$Hz) could significantly reduce the probability of
detection, even with GW templates that include spin effects
\cite{kidder95}. In the matched filtering technique, if the signal
and template are off by as little as half a wavelength (over
$\sim10^3$ cycles), the event could be missed because of
destructive interference \cite{owen}. In order to quantify this
threat, we attempt to measure Lyapunov divergence times for a
broad sample of initial conditions.

As in the Newtonian two-body problem, a relativistic binary system
can be expressed in terms of a reduced mass $\mu=m_1m_2/(m_1+m_2)$
orbiting a fixed mass $m=m_1+m_2$ with a separation
$\mathbf{\vec{r}} = \mathbf{\vec{r}_1} - \mathbf{\vec{r}_2}$. We
adopt units with $G=c=m=1$ and use the same notation and
post-Newtonian (PN) equations of motion as described by Kidder
\cite{kidder93}, expanded to second order (2PN) terms in $(v/c)^2$
and $Gm/r$, and including spin-orbit (SO) and spin-spin (SS)
effects as well as the 2.5PN radiation reaction (RR) term. The
individual spins precess because of frame-dragging and the
Lens-Thirring effect according to $\mathbf{\dot{\vec{S}}}_i =
\mathbf{\vec{\Omega}}_i\times \mathbf{\vec{S}}_i$. Here $i=1,2$,
$\mathbf{\vec{S}}_i$ is the spin of the compact object, and
$\mathbf{\vec{\Omega}}_i$ is a variable axis of precession as
defined in eqn.\ (2.4) of \cite{kidder95}. The magnitude of
$\mathbf{\vec{\Omega}}_i$ is the instantaneous spin precession
frequency. We also define a mass ratio $\beta \equiv m_2/m_1$, and
a spin-orbit misalignment angle $\theta_i$ for each object as the
angle between the spin vector $\mathbf{\vec{S}}_i$ and the
Newtonian angular momentum $\mathbf{\vec{L}}_N =
\mu(\mathbf{\vec{r}} \times \mathbf{\dot{\vec{r}}})$. The relative
position vector $\mathbf{\vec{r}}$ evolves according to a
second-order ordinary differential equation of the type
$\mathbf{\ddot{\vec{r}} =
\vec{a}_{PN}+\vec{a}_{SO}+\vec{a}_{SS}+\vec{a}_{RR}}$. The full
expressions for these terms can be found in Refs.~\cite{levingrqc}
and~\cite{kidder95}.

We integrate these equations numerically in double precision using
a 5th-order Runge-Kutta scheme with an adaptive time step. The
robust nature of the Runge-Kutta algorithm makes it particularly
attractive for measuring exponential divergence of nearby
trajectories. Indeed, since the Runge-Kutta integration can only
introduce errors that grow at a polynomial rate, any exponential
divergence, if it occurs, will rapidly dominate the evolution and
should be easily distinguishable from numerical effects
\cite{sussman}. To quantify chaotic behavior in a dynamical
system, the equations of motion must be conservative in the sense
that there should be no dissipative terms that could act as
attractors in phase space (which would eliminate the possibility
of the system being formally chaotic; see \cite{cornish}). For
this reason, when integrating the equations of motion to calculate
Lyapunov exponents, the radiation reaction terms are not included.
This also allows for the option of very long integrations to test
whether the system is formally chaotic (but on a time scale much
longer than the actual inspiral time).

Historically, an important tool for identifying chaos in celestial
mechanics has been the Poincar\'e surface of section
\cite{murray}. By plotting the position in phase space only at
certain values of independent coordinate variables, one can reduce
a complicated four-dimensional (4-D) phase space trajectory to a
2-D scatter plot. Conserved quantities such as energy or angular
momentum generally are held constant for different initial
conditions plotted on a single section. For many systems, some
initial conditions behave regularly, producing 1-D curves as if
there were additional integrals of the motion, while others fill
out 2-D regions of the section. This spreading away from the
invariant curves is evidence for chaotic behavior in the system.
However, this method is only successful in reducing the phase
space by one additional dimension, so for systems with [\# of
degrees of freedom]$-$[\# of integrals of motion] greater than
two, the surface of section technique is not very useful for
identifying chaos \cite{wisdom}. For higher-dimensional systems,
the projection of non-chaotic orbits onto a two-dimensional
section in general will not generate confined curves, and thus a
spreading of points in this pseudo-section is not necessarily an
indication of chaos. For the problem of two spinning compact
objects, the phase space is 10-D (3 from $\mathbf{\vec{r}}$, 3
from $\dot{\mathbf{\vec{r}}}$, and 2 each from the spin vectors
$\mathbf{\vec{S}_1}$ and $\mathbf{\vec{S}_2}$, which can precess
in arbitrary orientations but maintain constant magnitude
\cite{phase_dim}) while there are only 4 obvious constraints,
corresponding to angular momentum and energy conservation due to
the invariance of the PN Lagrangian under rotations and time
translations \cite{kidder93}. This almost guarantees that the
projection of the trajectory onto a 2-D section of phase space
will \textit{not} be constrained to a 1-D curve, so regular
behavior could easily be misinterpreted as chaos (cf.\
\cite{levinprl, levingrqc}).

A more quantitative method for identifying and measuring chaos is
to calculate the maximum Lyapunov exponent, defined as the
divergence rate between initially nearby trajectories:
\begin{equation}
\gamma(t) \equiv \frac{1}{t} \hspace{0.1cm} \mathrm{ln}
\left(\frac{dX(t)}{dX(0)}\right).
\end{equation}
Here the difference $dX$ between two points in phase space is
simply the Cartesian distance between the dimensionless
12-component coordinate vectors $[\mathbf{\vec{r}},
\dot{\mathbf{\vec{r}}}, \mathbf{\vec{S}_1}, \mathbf{\vec{S}_2}]$
and $[\mathbf{\vec{r}'}, \dot{\mathbf{\vec{r}'}},
\mathbf{\vec{S}_1}', \mathbf{\vec{S}_2}']$ of two nearby
trajectories. In our numerical integrations, the initial
separation is a small displacement in phase space with a random
orientation and a magnitude of $dX(0) = 10^{-10}$. The two
trajectories are then integrated forward in time, recording the
separation $dX(t)$, from which $\gamma$ is computed. In chaotic
systems, the divergence will be exponential in time with a roughly
constant (positive) exponent: $dX(t) \sim dX(0) \exp(\gamma t)$.
To be precise, there are actually many different Lyapunov
exponents, one for each dimension of phase space. The
\textit{maximum} Lyapunov exponent is automatically selected
because, like a classical eigenvector problem, any vector (such as
the random initial displacement vector) when multiplied repeatedly
by the same matrix will grow fastest in the direction
corresponding to the largest eigenvalue. Similarly, for a chaotic
system, any random initial displacement in phase space is expected
to grow as fast as the maximum Lyapunov exponent. However, for a
regular system, the divergence generally grows linearly or at most
as a power law in time. In that case, the Lyapunov exponent
$\gamma$ approaches zero for large $t$. We define the Lyapunov
time $t_{\rm L} \equiv 1/\gamma$ as the time scale on which nearby
trajectories separate by a factor of $e$ (the ``e-folding time'').
While it is difficult, on the basis of numerical integrations, to
claim that a system is categorically non-chaotic on all time
scales, we \textit{can} make the more precise claim that on a
particular time scale the system shows no chaotic behavior, that
is, we can set a lower limit on $t_{\rm L}$. For the problem of
coalescing compact binaries, the relevant time scale is the
inspiral time $t_{\rm insp}$, so if $t_{\rm insp} \ll t_{\rm L}$,
chaos will not affect the dynamics.

To establish the validity of our numerical approach, we have
re-examined the problem studied by Suzuki and Maeda \cite{suzuki},
where both chaotic and regular trajectories have clearly been
identified for super-maximally spinning test particles ($S_2 \gg
m_2^2$; $m \gg \mu$) orbiting around a Schwarzschild black hole
($S_1 = 0$). While physically unrealistic, these conditions
introduce no mathematical singularities into the equations of
motion and are therefore fine examples for identifying chaotic or
regular behavior in this dynamical system. Here we show numerical
results for two different sets of initial conditions: both begin
with $r/m = 4.5$, $\beta = 10^{-4}$, $S_1 = 0$, $S_2 = 1.4\times
10^4\ m_2^2$, and spin-orbit misalignment angle $\theta_2 =
\pi/4$, but one trajectory has an initial Newtonian orbital
angular momentum of $L_N/m = 1.485\ \mu$ while the other begins
with $L_N/m = 1.53\ \mu$. In Fig.~1 we show the measured value of
$\gamma$ as a function of time (measured in orbital periods). The
orbit with $L_N/m = 1.485\ \mu$ (a) is clearly chaotic with
$\gamma \sim 0.1$ (corresponding to a divergence time scale of
$t_{\rm L} \sim 10$ orbital periods). The other trajectory (b)
exhibits very different qualitative behavior, with no evidence for
chaos even on the time scale of thousands of orbits. Our results
for these and many other trajectories that we have computed agree
well with Suzuki and Maeda's results, confirming that both chaotic
and regular orbits exist for rapidly spinning test particles
around a Schwarzschild black hole.

Another method for distinguishing between chaotic and
quasi-periodic orbits is to look for certain qualitative features
in the power spectrum of one of the dynamical variables
\cite{sussman, sussman92}. For the compact binary problem, the
obvious choice is to analyze the spectrum of the GW signal,
defined as the squared amplitude of the Fourier transform of the
GW strain $h(t)$. With RR turned off, the spectrum should exhibit
sharp lines for quasi-periodic orbits, but broad-band noise for
chaotic orbits. Including only the leading quadrupole radiation
terms, the components of the transverse-traceless GW tensor
observed from the $\hat{z}$ direction are
\begin{equation}
h_+ = \frac{4\mu}{D} \left(v_x^2 - \frac{m}{r^3}x^2 \right),
\end{equation}
\begin{equation}
h_\times = \frac{4\mu}{D} \left(v_x v_y - \frac{m}{r^3}xy \right),
\end{equation}
where $D$ is the distance to the source and $v_i = \dot{x}_i$
\cite{kidder95, mtw}. Figure 2 shows the power spectra in $h_+(t)$
for the same two test-particle orbits as in Fig.~1. Both spectra
show peaks around $f_{\rm GW} \simeq 25\,$Hz, the fundamental
quadrupole frequency (twice the orbital frequency) for these
initial conditions. As expected, the chaotic orbit (a) produces
primarily broad-band noise with few distinguishable features,
whereas the spectrum for the quasi-periodic orbit (b) shows many
sharp lines.

Having established the ability of our method to distinguish
between regular and chaotic trajectories described by the PN
equations of motion, we now apply the same techniques to the
astrophysically relevant systems expected to be detectable by
ground-based laser interferometers. Here we will show results for
an illustrative case of two maximally spinning $10M_\odot$ black
holes ($S_i = m_i^2$) in an initially circular orbit with
spin-orbit misalignment angles of $\theta_1 = 38^o$ and $\theta_2
= 70^o$. To measure the Lyapunov exponent at different points
during the final inspiral, 4 different trajectories were
integrated, corresponding to GW frequencies of 10, 40, 100, and
400 Hz, emitted at orbital separations $r/m\simeq$ 50, 20, 10, and
5, respectively \cite{initnote}. For each separation, we then
integrate the PN equations of motion (with RR turned off) over a
period much longer than the inspiral time. If no exponential
divergence is observed, we can safely conclude that the system is,
for all practical purposes, not chaotic.

The measured values for $\gamma(t)$ are plotted in Fig.~3 for
each of the selected stages of the inspiral. For any power-law
divergence in phase space with $dX(t) \sim t^\alpha$, on a log-log
plot of $\gamma$ versus time the slope is $d \ln \gamma/ d \ln t =
1/\ln t -1$, so that for large times the curve should be linear
with slope -1 [as we see in Fig.~1 (b)]. All plots
in Fig.~3 clearly show this behavior, characteristic of regular,
non-chaotic orbits. Also shown for comparison in Fig.~3 are the
inspiral times corresponding to each separation. In all four
cases, the lower limit on $t_{\rm L}$ far exceeds $t_{\rm insp}$.

Just as in the test particle example presented above, we also
calculate the gravitational wave power spectrum for this black hole
binary. Here the results are even more striking: Fig.~4 shows
the $h_+(t)$ power spectrum at the point in the inspiral where the
GW frequency is $100\,$Hz. The spectrum has a few very sharp
lines with no evidence for broad-band noise. The strongest
features correspond to the fundamental GW quadrupole and spin-orbit
precessional frequencies ($100\,$Hz and $7\,$Hz, respectively) and their
harmonics. The spectra from the other points in the inspiral show
the same qualitative features, all indicative of quasi-periodic
motion.

In addition to this prototypical binary black hole system, we have
investigated the behavior of a large number of other systems at
different stages throughout the inspiral. Varying the binary mass
ratio, spin magnitudes (always with $S_i \leq m_i^2$),
misalignment angles, eccentricity, and initial separation over
wide ranges, we consistently find the same regular, non-chaotic
behavior for all trajectories. In particular, we have calculated
orbits for the same high-eccentricity systems considered by Levin
in Ref.~\cite{levinprl} and again measure only linear divergence
of nearby initial conditions, finding $t_{\rm insp} \ll t_{\rm L}$
\cite{belczynski}. We have also looked at many binary systems in
which either one or both objects are solar-mass neutron stars.
Because of their smaller mass (and thus slower radiation loss),
these systems can spend significantly longer times in the inspiral
band, completing tens of thousands of orbits before merger. Yet,
even on the longest relevant time scales ($t_{\rm insp} \gtrsim
10^4\,$s), we again find no chaotic behavior.

Recent work in binary star evolution \cite{kalogera} suggests that
systems with large spin misalignment angles are likely to be
astrophysically relevant sources, giving rise to complicated
nonlinear SO and SS interactions. It is possible that there are
undiscovered narrow regions in phase space where specific resonant
conditions could give rise to chaotic behavior, yet even if a
binary system passes through them during inspiral, the time spent
in these resonant bands will most likely be much shorter than the
Lyapunov time. Even though the Lyapunov exponent approaches zero
for all the systems we have considered, their orbits can still
appear quite irregular, especially when both objects are spinning
rapidly \cite{levinprl}. As mentioned above, this can lead to the
spreading of the surface of section away from a 1-D curve, but in
dynamical systems with many degrees of freedom, this spreading
alone does not imply the presence of chaos. Nevertheless, the
detection of GW signals from binaries with rapidly spinning
components may be very challenging. Systems with similar initial
conditions still may produce waveforms that look quite different
even without exponential divergence. Orbital precession will add
many complicated features to the basic inspiral waveforms. The
current template database will most likely need to be extended
(perhaps by orders of magnitude) to include at least a portion of
parameter space where the signals are significantly modified by
spin effects \cite{apost}. Furthermore, for the late stages of
inspiral, the 2.5PN expansion becomes invalid and one should
include higher order PN terms (or even perform calculations in
full general relativity \cite{buonanno}). While these terms
introduce additional features in the dynamics, at this late stage
the orbit is certainly decaying too rapidly for the final GW
signal to be affected by formal chaotic behavior.

\begin{acknowledgments}
We are very grateful to J.\ Wisdom for many useful comments and
discussions. We also thank V.\ Kalogera and J.\ Levin for helpful
discussions. This work was supported by NSF Grants AST-9618116 and
PHY-0070918 and the NSF Graduate Research Fellowship Program.
\end{acknowledgments}

%Figure 1.
\begin{figure}[bp]
\includegraphics{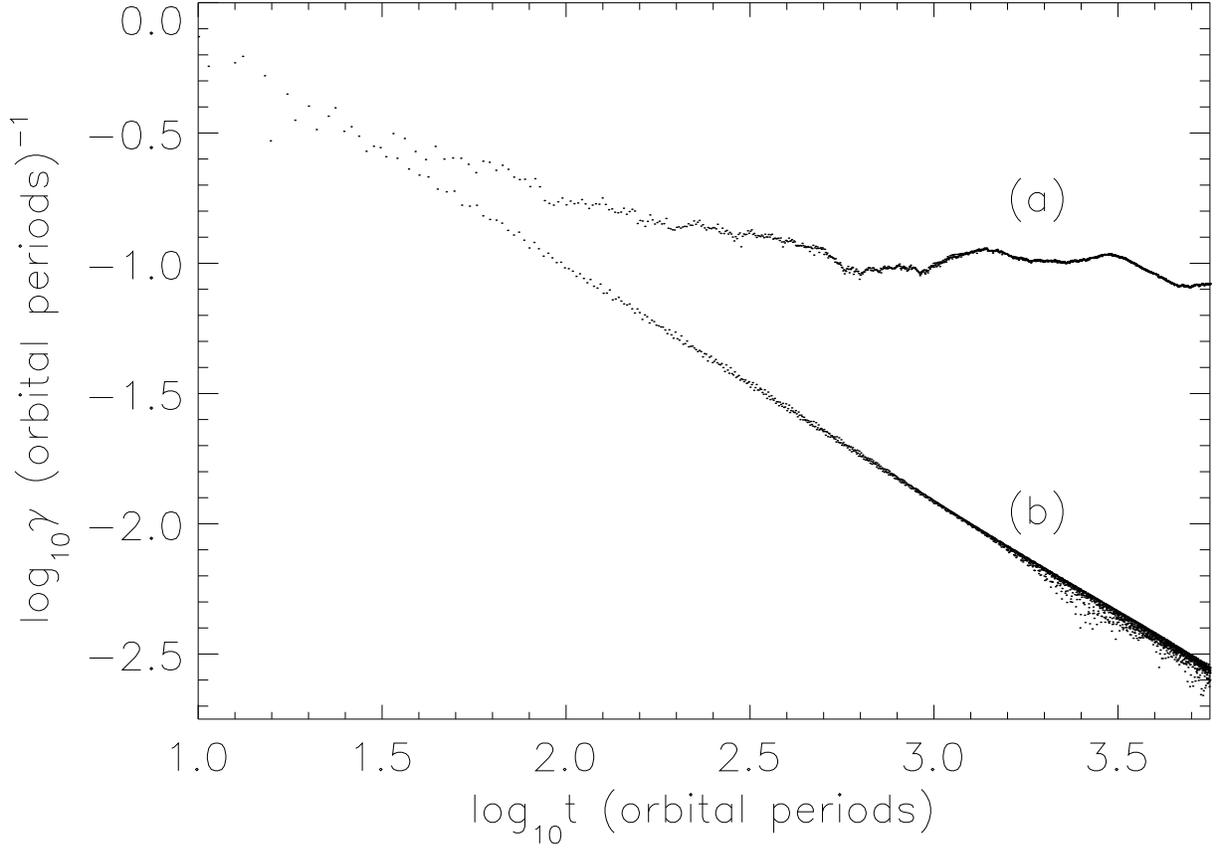}
\caption{Lyapunov exponent $\gamma$ for test-particle orbits
around a black hole. The orbits in case~(a) are chaotic, with
$\gamma$ approaching a positive value corresponding to a
divergence time of $t_{\rm L} \sim 10$ orbital periods. The
regular orbits in case~(b) diverge linearly in time so that
$\gamma(t) \to 0$.}
\end{figure}

%Figure 2.
\begin{figure}[bp]
\includegraphics{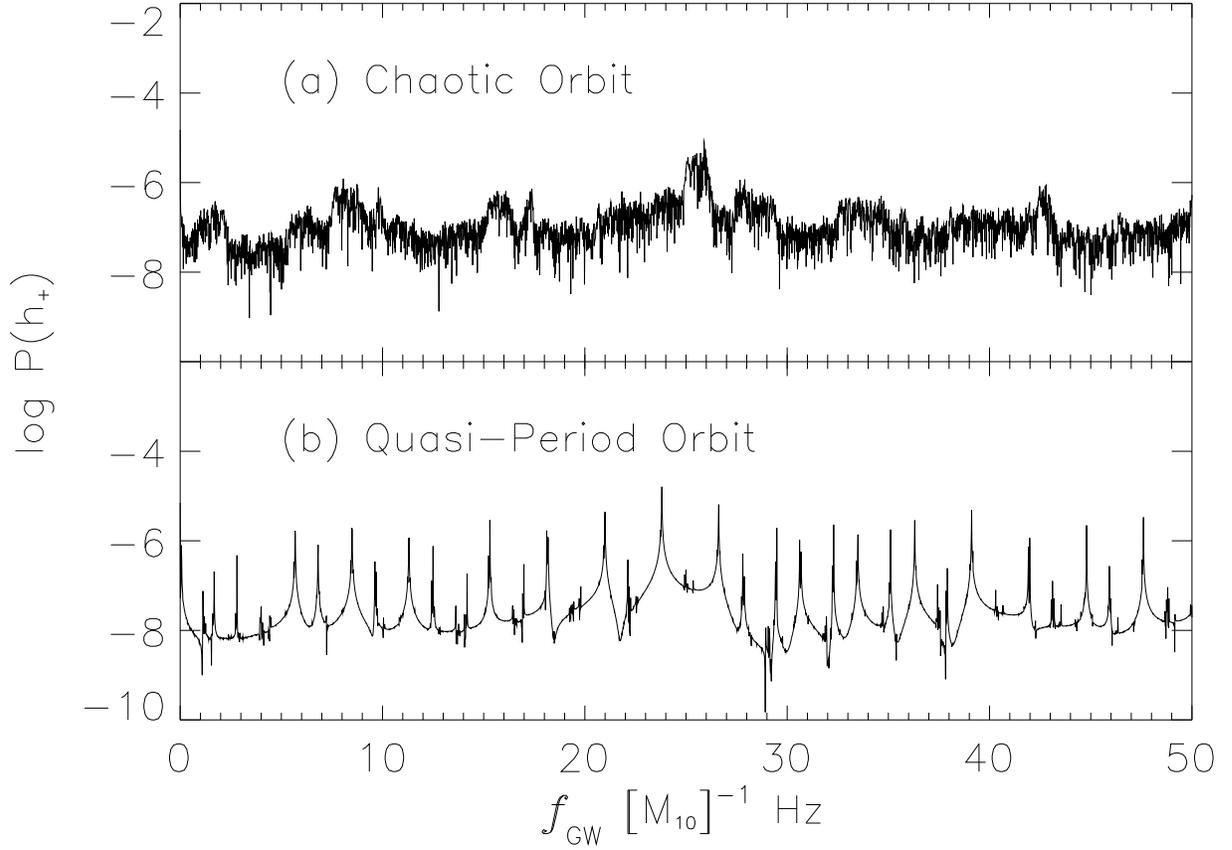}
\caption{Power spectra of the GW signal $h_+(t)$ for the same two
cases as in Fig.~1. The chaotic system in (a) produces broadband
noise while the quasi-periodic orbit (b) exhibits sharp spectral
lines. The GW frequency is given in units of $m_{10}^{-1}\,$Hz,
where $m_{10}=m/10M_\odot$.}
\end{figure}

%Figure 3.
\begin{figure}[bp]
\includegraphics{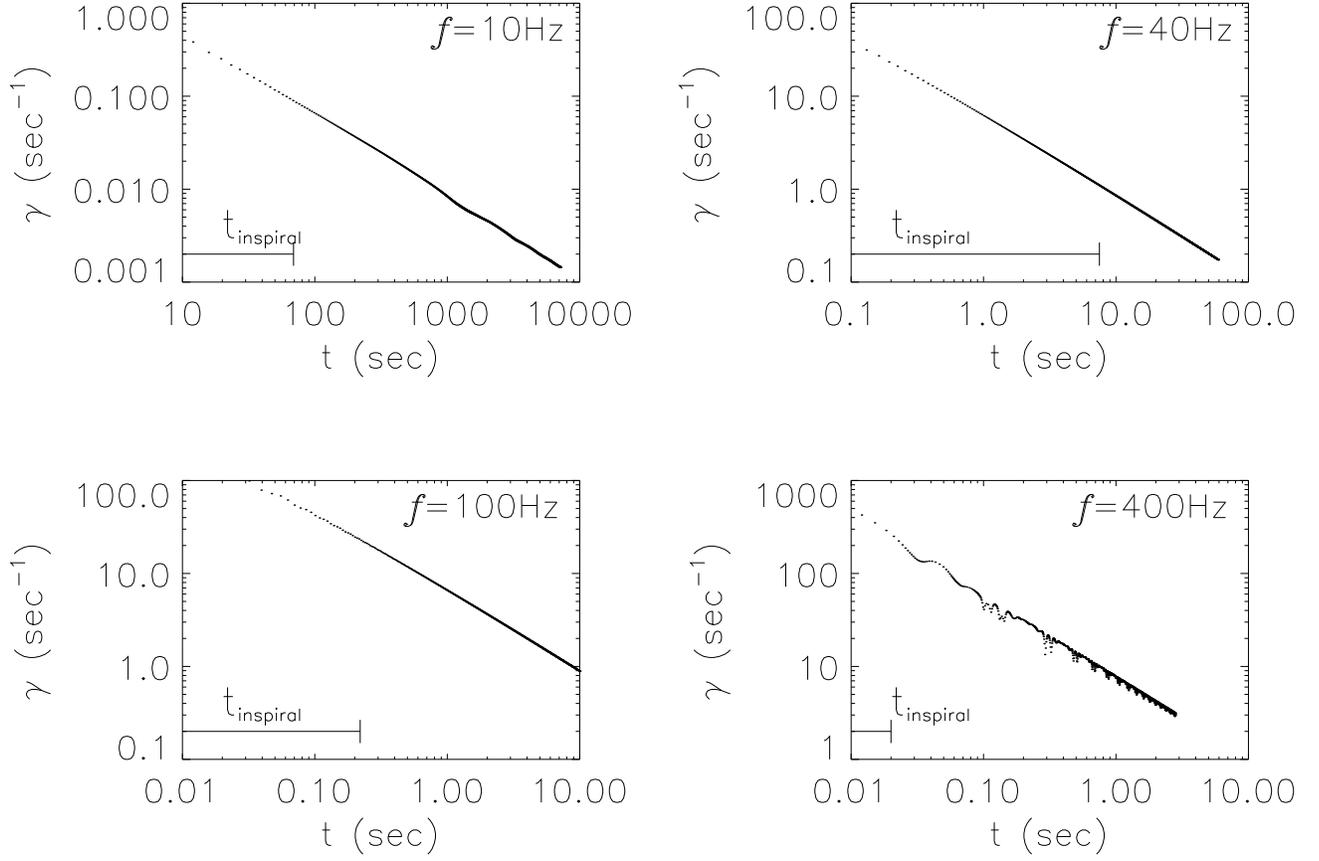}
\caption{Lyapunov exponent $\gamma(t)$ for different stages in the
inspiral of two maximally spinning $10M_\odot$ black holes:
$f_{\rm GW} = 10\,$Hz, $r/m=47.25$, $t_{\rm insp} = 69\,$s;
$f_{\rm GW} = 40\,$Hz, $r/m=18.75$, $t_{\rm insp} = 7.5\,$s;
$f_{\rm GW} = 100\,$Hz, $r/m=9.2$, $t_{\rm insp} = 0.2\,$s;
$f_{\rm GW} = 400\,$Hz, $r/m=4.0$, $t_{\rm insp} = 0.01\,$s. No
evidence for chaos is seen, with $t_{\rm L}=1/\gamma \gg t_{\rm
insp}$ in all cases.}
\end{figure}

%Figure 4.
\begin{figure}[bp]
\includegraphics{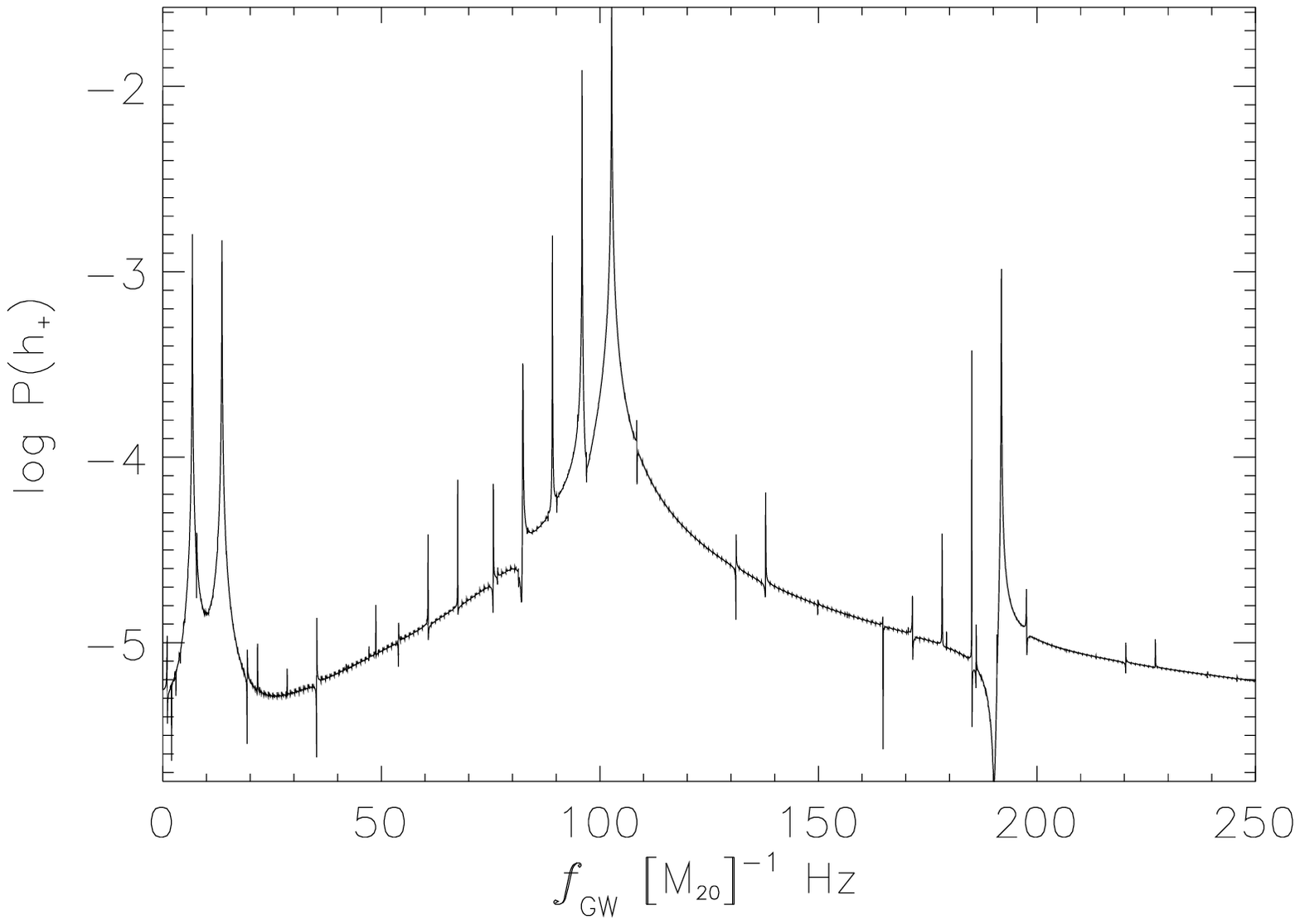}
\caption{Power spectrum of the GW signal $h_+(t)$ calculated from
the initial conditions of Fig.~3 at the stage in the inspiral
where $f_{\rm GW} = 100\,$Hz. The frequency is in units off
$m_{20}^{-1}\,$Hz, where $m_{20}=m/20M_\odot$. The sharp lines in
the spectrum confirm that the orbit is regular.}
\end{figure}


\begin{thebibliography}{99}
\bibitem{phase_dim} We do not consider the spin momenta as
dimensions in phase space since they play no role in the equations
of motion. The precession equations in \cite{kidder95}
independently give the instantaneous spin momenta in terms of
other phase space variables.
\bibitem{initnote} These four different trajectories are actually all
computed from the same initial conditions at $r/m = 47.25$,
turning on RR to evolve the system from one stage to the next.
\end{thebibliography}
\end{document}